\documentstyle[epsf]{l-aa} 
\begin{document}

\newcommand{\Kkms}{\,{\rm K\,km/s} }
\newcommand{\WHI}{\,{W_{\rm HI}} } 

\newcommand{\mic}{\,{\rm \mu m} } 
\newcommand{\pac}{\,{\rm pc} } 

\newcommand{\Wmsr}{\,{\rm W/m^2/sr} }
\newcommand{\Wcmsr}{\,{\rm W/cm^2/sr} }
\newcommand{\Wcm}{\,{\rm W/cm^2} }
\newcommand{\cmun}{\,{\rm cm^{-1}} }
\newcommand{\Mj}{\,{\rm MJy/sr} }

\thesaurus{09.07.1,     
	   09.04.1,     
	   12.04.2, 	
           13.09.3,     
           13.18.3,	
           13.09.2}	

\title{First detection of the Warm Ionised Medium dust emission.}
\subtitle{Implication for the Cosmic Far-Infrared Background}

    \author{G. Lagache  
          \inst{1} \and 
          A. Abergel 
          \inst{1} \and 
          F. Boulanger 
          \inst{1} \and
          F.X. D\'esert
          \inst{2} \and 
           J.-L. Puget 
          \inst{1}}

 \offprints{G. Lagache } 
 
\institute{$^1$ Institut d'Astrophysique Spatiale, B\^at.  121, 
Universit\'e Paris XI, F-91405 Orsay Cedex\\
$^2$ Laboratoire d'Astrophysique, Observatoire de Grenoble,
BP 53, 414 rue de la piscine, F-38041 Grenoble Cedex 9}

 \date{Received 15 July 1998, Accepted 29 september 1998}  
 
   \maketitle
\markboth{First detection of the WIM dust emission. Implication for
the CFIRB}{}

\begin{abstract}
We present a new analysis of the far-IR emission 
at high Galactic latitude based on COBE and HI data.
A decomposition of the Far-IR emission over the HI, H$^{+}$ and H$_2$
Galactic gas components and the Cosmic Far InfraRed Background (CFIRB)
is described. \\

For the first time the far-IR emission
of dust associated with the Warm Ionised Medium (WIM) is evidenced.
This component determined on about 25$\%$ of the sky
is detected at a 10$\sigma$ level in the [200, 350]~$\mic$ band. 
The best representation of the WIM dust spectrum is
obtained for a temperature of 29.1 K and an emissivity law
$\tau/N_H^+=$~3.8~$\pm$~0.8~$10^{-26}~(\lambda/250\mic)^{-1}$~cm$^2$.
With a spectral index equal to 2, the emissivity law becomes
$\tau/N_H^+=$~1.0~$\pm$~0.2~$10^{-25}~(\lambda/250\mic)^{-2}$~cm$^2$,
with a temperature of 20~K, which is significantly higher
than the temperature of dust associated with HI gas.
The variation in the dust spectrum from the HI to the WIM component
can be explained by only changing the upper cutoff
of the Big Grain size distribution from 0.1~$\mic$ to 30 nm.\\
The detection of IR emission of dust in the WIM significantly 
decreases the intensity of the CFIRB, especially
around 200~$\mic$ which corresponds to the peak of energy. \\

\end{abstract}

\section{Introduction}
The extraction of the Cosmic Far Infrared Background (CFIRB), induced by the emission of
light from distant galaxies (Partridge \& Peebles, 1967; Bond et al., 1986 and
references therein), requires an accurate subtraction of the Interstellar Medium (ISM) 
foreground emissions. The two instruments DIRBE and FIRAS on board the COBE satellite
provide actually the best available data to study, on the whole sky, the distribution and
properties of the ISM far InfraRed (far-IR) emission. \\

Boulanger et al. (1996) have extensively studied the emission of the dust associated
with the HI component using the spatial correlation between the 
far-IR dust emission as measured by DIRBE and FIRAS and the 21 cm HI
emission as measured by the Leiden/Dwingeloo survey of the northern hemisphere.
The dust emission spectrum derived from this correlation (for 
N$_{HI}<$~4.5~10$^{20}$~cm$^{-2}$) can be quite well represented by a 
single modified Planck curve characterized by T=17.5~K and  
$\tau/N_{HI}~=~10^{-25}~(\lambda/250\mic)^{-2}$~cm$^2$. 
This emissivity law is very close 
to the one predicted by the Draine $\&$ Lee (1984) dust model. \\

Dust emission associated with molecular clouds has been recently studied 
through Far-IR and submillimeter (submm) observations with the DIRBE, FIRAS and SPM/PRONAOS  
instruments. In a previous paper (Lagache et al., 1998), 
we have extensively studied the spatial distribution
of the temperature of the dust at thermal equilibrium using the DIRBE and
FIRAS experiment. We have 
found at large angular scale the presence of a cold dust component 
(with a median temperature of 15 K), very well correlated
with molecular complexes with low star forming activity such as Taurus.
The lowest values of the temperature found in the cold regions ($\sim13$~K) are 
comparable with that obtained for dense clouds in star forming regions by the 
balloon-borne experiment SPM/PRONAOS (Ristorcelli
et al., 1996, 1998, Serra et al., 1997).
The association between the cold dust component and molecular clouds is further demonstrated
by the fact that all sky pixels with significant cold emission have an 
excess IR emission with respect to the high latitude IR/HI correlation.
A threshold value of the column density, 
N$_{HI}$=2.5~$10^{20}$~H~cm$^{-2}$, below which cold dust 
is not detected within the FIRAS beam of $\sim7\degr$ has been deduced.
This knowledge on the spatial distribution of the dust associated with cold molecular
clouds is important for the search of the CFIRB since it allows to select 
parts of the sky for which cold dust is not detected.\\

On the other hand, 
the knowledge of the dust emission associated with the H$^{+}$ component is very poor.
Observations of H$_{\alpha}$ emission at high Galactic latitudes 
and of dispersion measures in the 
direction of pulsars at high $|z|$ indicate that the low-density ionised gas (the Warm 
Interstellar Medium, WIM) accounts for some 30$\%$ of the gas in the solar neighborhood
(Reynolds, 1989). There is also evidence that part of the
WIM is spatially correlated with the HI gas (Reynolds et al., 1995). 
Consequently, a significant fraction of the Far-IR emission
associated with the WIM may contribute to the spectrum of the dust associated
with the HI gas.
However, the scale height of the H$^{+}$ medium is much larger than the HI one, so
a significant part of the H$^{+}$ is completely uncorrelated with the HI.
Since most of the
grain destruction is expected to occur in the low-density component of the ISM (Mc Kee 1989),
the WIM could also be dust poor. Depletion studies of elements that form the grains 
show that grains are indeed partly destroyed in the low
density phases of the ISM (review by Savage \& Sembach, 1996).
Measuring the dust emission from the WIM 
could allow to
understand the evolution of the dust in the low-density gas. 
However, this measure is difficult
because one can not easily separate the contribution of the H$^{+}$ gas from that 
of the HI.
Boulanger \& Perault (1988)
unsuccessfully searched in the 100~$\mic$ IRAS all-sky map for such a contribution.
The unfruitful analysis may be due to the spatial correlation between
the HI and H$^{+}$ emissions. 
Boulanger et al. (1996) have searched
for such a component in the residual FIRAS emission after the removal of the HI component.
They found that the residual emission is consistent with an emission spectrum 
like that of the
HI gas for N$_{HI}\sim$~4~10$^{19}$~cm$^{-2}$. However,
they consider this as an upper limit
for the contribution of the H$^{+}$ component since they could have measured emission
from low contrasted molecular clouds. Arendt et al. (1998) have also investigated potential
IR WIM dust emission. They conclude that they were unable to
detect any IR emission associated with low density ionised gas at high Galactic latitude
(the fraction of the sky used is less than 0.3$\%$). However, very recently,
Howk \& Savage (1999) have pointed out, for the first time, the existence
of Al- and Fe-bearing dust grains towards two high-z stars. They have shown
that the degree of grain destruction in the ionised medium, through
these two stars, is not much higher than in the warm neutral medium. If dust
is present in the WIM, one should detect its infrared emission.\\

The CFIRB is expected
to have two components: direct starlight redshifted in the far-IR and submm, 
and the stellar radiation absorbed by dust.
We concentrate here on the submm part
of this background. Its detection is very difficult because of the 
strong and fluctuating Galactic foregrounds. 
First, upper limits have been
reported: Hauser et al. (1995) from DIRBE data and
Mather et al. (1994) from FIRAS data. Lower limits on the CFIRB have been obtained from the
deepest IRAS and K galaxy counts (Hauser et al., 1994 and references therein). 
The first direct detection of the CFIRB has been reported
by Puget et al. (1996). All the Galactic foregrounds were modeled and removed using independant
dataset in addition to the FIRAS data. Its spectrum indicates the presence of sources at
large redshift. The main uncertainty on the CFIRB 
comes from Galactic foregrounds. Therefore, we stress that the Puget
et al. (1996) results were confirmed in the cleanest parts 
(N$_{HI}<10^{20}$~cm$^{-2}$ in a 7$\degr$ beam) of the sky (Guiderdoni et al.,
1997). More recently, Fixsen et al. (1998) and Hauser et al. (1998) have  
also confirmed the detection of the CFIRB using FIRAS and DIRBE data.\\

A general problem with all these determinations of the CFIRB comes from
a potential Far-IR emission from the WIM which has never
been determined. The goal of this paper is to push our knowledge
of the Galactic emission one step forward by deriving the far-IR spectrum of the
WIM dust emission. Then, we use our understanding of the
interstellar dust emissions associated with the HI and H$^{+}$ components 
to give a more accurate estimate of the CFIRB spectrum. The paper is
organised as follow:
in Sect. \ref{sect_data_prep}, we present the data we have used.
The variations of the dust emission spectrum associated with different 
HI gas column densities are studied in Sect. \ref{sect_hi_var}. In Sect. 
\ref{sect_effect_hii}, we show that the spatial variations of the
dust emission spectrum in the low HI column density regions
can be due to the presence
of the non-correlated H$^{+}$ component. After the removal of the dust HI component, we
detect a residual Galactic emission which is attributed to the WIM (Sect. \ref{sect_hii_spec}).
This is the first detection of the WIM dust emission. 
We show (Sect. \ref{sect_CFIRB_LH}) that the FIRAS spectra in the very low HI
column density regions
exhibit a large excess over the emission of dust associated with HI and H$^{+}$ components.
In these
regions, the CFIBR dominates the FIRAS emission.
We test its isotropy in Sect. \ref{sect_CFIRB_iso}. 
All the results are summarised in Sect. \ref{sect_cl}.  

\section{\label{sect_data_prep} Data presentation and preparation}
The FIRAS instrument is a polarising Michelson 
interferometer with $7\degr$ resolution and two separate bands which have a
fixed spectral resolution of 0.57~cm$^{-1}$ (Fixsen et al. 1994). The low frequency band 
(2.2 to 20~cm$^{-1})$ was designed to study the 
CMB (Cosmic Microwave Background) and the high frequency band (20 to 96 cm$^{-1}$) 
to measure the 
dust emission spectrum in the Galaxy. We use the 
so-called LLSS (Left Low Short Slow) and RHSS (Right High Short Slow) data from the 
"pass 3" release which cover the low and high frequency bands respectively
(see the FIRAS explanatory supplement).\\
DIRBE is a photometer with ten bands covering the range
from 1.25 to 240$~\mic$ with 40 arcmin resolution (Silverberg et al. 1993). 
We choose to use annual 
averaged maps because they have a higher signal to noise ratio than maps
interpolated at the solar elongation of $90\degr$  
(see the DIRBE explanatory supplement). In our analysis, we only use the 140 and
240~$\mic$ maps. The present study is based on "pass 2" data. \\
Since in our analysis we combine the FIRAS and DIRBE data at 140 and 240~$\mic$,
we convolve the DIRBE maps with the FIRAS Point Spread Function 
(PSF). 
The PSF is not precisely known for all wavelengths, so we use the approximation 
suggested by Mather (private communication) 
of a $7\degr$ diameter circle convolved with a line of $3\degr$ length
perpendicular to the ecliptic plane (Mather et al. 1986).\\
Before studying the Far-IR emission, we have subtracted the CMB and its dipole 
emission from the FIRAS data using the parameters given by 
Mather et al. (1994) and Fixsen et al. (1994).
To remove the InterPlanetary (IP) dust emission, we first consider the $25~\mic$ map as
a spatial template for the IP dust. Then, we compute the IP dust
emission at 100~$\mic$ using the zodiacal emission ratio given by Boulanger et 
al. (1996): $I_{\nu}(100)/I_{\nu}(25)=0.167$. We remove the IP dust emission at 
$\lambda\ge140\mic$ using the IP emission template at 100$\mic$ and considering a 
zodiacal spectrum $I_{\nu}\propto\nu ^3$ (Reach et al., 1995).\\ 

The HI data we used are those of the Leiden/Dwingeloo survey, which covers the
entire sky down to $\delta=-30\degr$ with a grid spacing of 30' in both l and
b (Hartmann $\&$ Burton, 1997). 
The 36' Half Power Beam Width (HPBW)
of the Dwingeloo 25-m telescope provides 21-cm maps at an angular resolution
which closely matches that of the DIRBE maps. These data represent an improvement over earlier
large scale surveys by an order of magnitude or more in at least one of the principal 
parameters of sensitivity, spatial coverage, or spectral resolution. Details of the
observationnal and correction procedures are given by Hartmann (1994) and by 
Hartmann $\&$ Burton (1997). The 21cm-HI data are convolved 
with the FIRAS PSF. We derive the HI column densities
with
1~K~km~s$^{-1}$=1.82~$10^{18}$~H~cm$^{-2}$ (optically thin emission).\\

Throughout this paper, diffuse parts of the sky are selected following
Lagache et al. (1998). To remove 
molecular clouds and HII regions, we use the DIRBE map of the
240~$\mic$ excess with respect to the 60$\mic$ emission:
$\Delta$S=S$_{\nu}$(240)-~C$\times$S$_{\nu}$(60) with C=4$\pm$0.7.
This map shows as positive flux regions, the cold component of the dust emission,
and as negative flux regions, regions where the dust is locally
heated by nearby stars (like the HII regions).
Therefore, diffuse emission pixels are selected with $|\Delta S| < x \sigma$, 
$\sigma$ being evaluated from the width
of the histogram of $\Delta$S and $x$ being chosen for our different purposes.
For example, in Sect. \ref{sect_hii_spec}, we take $x=1$,
which is very restrictive,
to ensure that the selected pixels are mostly coming from the diffuse medium;
in Sect. \ref{sect_CFIRB_iso}, we take $x=3$, since we need
a fraction of the sky as large as possible. \\
We also use the 240~$\mic$/HI map excess of Reach et al. (1998). Regions
for which this excess is greater than 3$\sigma$ are systematically discarded. \\
From Lagache et al. (1998), we have for each line of sight the spectrum of
the cold component of the dust emission (if cold dust is detected) and the cirrus spectrum.
These spectra are used to compute the contribution of the cold dust emission
in Sect. \ref{sect_hi_var} (Table \ref{tbl-1}).

\section{\label{sect_hi_var} Emission spectrum of dust associated with HI gas}

In this part, we first concentrate on the spatial variations of the dust emission spectrum
with the HI gas column densities. We deduce a column density threshold 
above which the contribution
of the cold dust component induces a significant submm excess with respect to the 
$\nu^2$ emissivity law. Then, we investigate the spectrum of the
dust associated with regions containing
HI column densities below this threshold.\\ 

\subsection{\label{sect_HI_var} Variation of the dust emission spectrum with the HI column densities}

\begin{table*}
\caption{Mean N$_{HI}$ computed for different sets of pixels (see Sect.\ref{sect_HI_var}).
Optical depths ($\tau$) and temperatures (T) of the HI dust emission spectra (derived following
Eq. \ref{eq_spec}). 
Residual flux integrated in the [609-981]~$\mic$ band are obtained by 
removing a single component $\nu^2$ modified Planck curve (defined by $\tau$ and T)
from each HI dust spectrum. The fifth column gives the percentage, in flux
in the [609-981]~$\mic$ band, of the detected cold emission with respect to 
the cirrus emission for the corresponding set
of pixels
(see Lagache et al., 1998 for more details).}
\label{tbl-1}
\begin{center} 
\begin{tabular}{|c|c|c|c|c|l|} \hline
Mean N$_{HI}$ & $10^{-6} \times \tau$ & T (K) &  Residual flux & $\%$ of detected \\
(H cm$^{-2}$) & (normalized at 250 $\mic$) & & 
($10^{-10}$ W m$^{-2}$ sr$^{-1}$) & cold emission \\ 
& & & for $N_{HI}=10^{20}$ H cm$^{-2}$ & \\ \hline
1.8 $10^{20}$ & 6.8 $\pm$ 1.8 & 18.0$\pm$1.2 & -1.0 $\pm$ 8.6 & 0.0 \\ 
3.2 $10^{20}$ & 9.8 $\pm$ 1.4 & 17.3$\pm$0.6 & 3.0 $\pm$3.7 & 0.0\\ 
4.1 $10^{20}$ & 9.7 $\pm$ 1.1 & 17.7$\pm$0.5 & 4.0 $\pm$ 2.7 & 0.9 \\ 
5.0 $10^{20}$ & 10.2 $\pm$ 1.0 & 17.6$\pm$0.4 & 4.4 $\pm$ 2.2 & 0.6 \\ 
6.3 $10^{20}$ & 11.3 $\pm$ 0.9 & 17.5$\pm$0.3 & 6.4 $\pm$ 1.6 & 3.4 \\ 
8.1 $10^{20}$ & 12.1 $\pm$ 0.8 & 17.8$\pm$0.3 & 9.5 $\pm$ 1.3 & 6.6 \\ 
9.9 $10^{20}$ & 13.4 $\pm$ 0.7 & 17.9$\pm$0.2 & 10.9 $\pm$ 1.1 & 7.1 \\ 
1.2 $10^{21}$ & 17.1 $\pm$ 0.6 & 18.0$\pm$0.2 & 13.1 $\pm$ 1.0 & 11.9 \\ 
1.4 $10^{21}$ & 16.3 $\pm$ 0.6 & 18.0$\pm$0.2 & 13.9 $\pm$ 0.8 & 12.3 \\
1.8 $10^{21}$ & 17.1 $\pm$ 0.6 & 17.6$\pm$0.2 & 13.9 $\pm$ 0.7 & 8.0 \\ \hline
\end{tabular}\\
\end{center}
\end{table*}

To compute the emission spectrum of dust associated with 
HI gas from low to large column densities, we use the same method as in 
Boulanger et al. (1996). We first select sky pixels according to their
HI column density
and sort them into 10 sets of pixels bracketed by the following values of
$N_{HI}$: 
[1.1, 2.7, 3.7, 4.6, 5.5, 7.3, 9.1, 10.9, 12.7, 16.4, 20.1]$\times$~10$^{20}$~H~cm$^{-2}$
which correspond to 
$W_{HI}$:
[60, 150, 200, 250, 400, 500, 600, 700, 900, 1100]~K~km~s$^{-1}$.
These different sets cover between 1.2$\%$ (for the highest column density) and 
16$\%$ (for the lowest) of the sky. \\
Dust emission spectra are computed for each set k using the equation:
\begin{equation}
\frac{dF}{dN_{HI}}(k,l) = \frac{<F>_k - <F>_l}{<N_{HI}>_k - <N_{HI}>_l}
\label{eq_spec}
\end{equation}
where $<F>_{i}$ corresponds to the mean FIRAS spectra computed for
the set of pixels i, and  $<N_{HI}>_{i}$ the mean HI column density for
the same set of pixels. Forming the difference in Eq. \ref{eq_spec}
removes, within statistical variance, any residual IR emission 
which is not correlated with the HI gas such as an
isotropic component.\\
In this part, $<F>_k$ correspond to the FIRAS spectra computed for the
10 sets of pixels given above and
$<F>_l$ to the spectrum derived in the very low HI column density
regions, $N_{HI}~\le~1.1$~$10^{20}$~H~cm$^{-2}$ (representing $\sim2\%$ of the sky). \\

For each set of pixels k, 
the dust spectrum $\frac{dF}{dN_{HI}}(k)$
is fitted by a modified Planck curve:\\
$\nu I_{\nu}=~\tau~\left(~\frac{\lambda}{250~\mic}~\right)^{-2}~B_{\nu}(T)$, \\
where
$\tau$ is the optical depth at 250~$\mic$, $B_{\nu}(T)$ is the
Planck curve (in W/m$^2$/sr) and $\lambda$ the wavelength in $\mic$.
The $\alpha$=2 emissivity index corresponds to standard interstellar
grains (Draine \& Lee, 1984). 
The mean temperature is equal to 17.7~K 
with no significant variations from set to set (Table \ref{tbl-1}, column 3). 
The optical depth ($\tau$) increases 
linearly with the HI column density up to N$_{HI}$=$10^{21}$~H~cm$^{-2}$. Then, it levels off
at $\sim$1.7~$10^{-5}$ (Table \ref{tbl-1}, column 2).\\

\begin{figure}  
\epsfxsize=8.cm
\epsfysize=7.cm
\hspace{2.cm}
\vspace{0.0cm}
\epsfbox{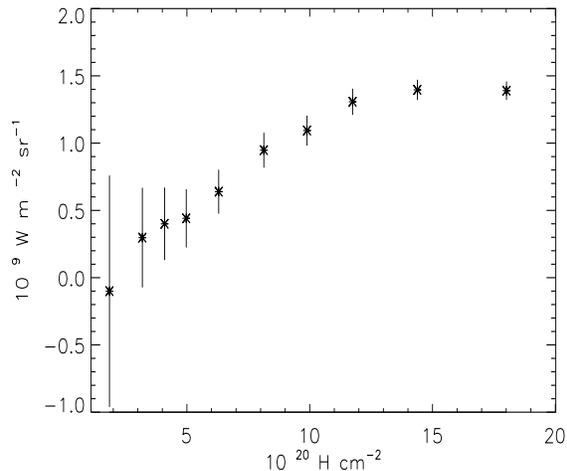}
\caption{\label{fig_res_hi} Residual flux 
integrated in the [609-981] $\mic$ band 
after removing a single component modified Planck curve with $\alpha$=2 
from dust spectra normalised at $N_{HI}$=10$^{20}$~H~cm$^{-2}$ and
computed for different HI column densities.} 
\end{figure}

We compute the residual spectrum for each set: \\
$\nu~R_{\nu}=~\frac{dF}{dN_{HI}}-~\nu I_{\nu}$,\\ and
study it by integrating $\nu R_{\nu}$ in the [609-981] $\mic$ 
band. Results are presented column 4 of Table 1 and Fig. \ref{fig_res_hi}. 
For N$_{HI}$ larger than 5~$10^{20}$~H~cm$^{-2}$, we detect a significant residual emission
($>3\sigma$). 
Spectra at high column density contain a positive submm 
excess with respect to a single temperature $\nu^{2}$ emissivity 
law Planck curve for $\lambda>600\mic$. 
In Lagache et al. (1998), we have shown, using DIRBE data (at the FIRAS resolution), 
that cold emission from interstellar dust associated 
with H$_{2}$ gas is only detected where
HI column densities are larger than $\sim$~2.5~10$^{20}$~H~cm$^{-2}$. 
In the Galaxy, at $|b|>10\degr$, 90$\%$ of this cold emission  
is distributed over regions with HI column densities larger than 5~$10^{20}$~H~cm$^{-2}$.
Column 5 of Table 1 gives the percentage, in flux
in the [609-981]~$\mic$ band, 
of detected cold emission for the different sets, derived from the analysis
of Lagache et al. (1998). 
We clearly see that the submm excess increases with the contribution of the cold 
emission.
Therefore, we conclude that this positive excess is due to the emission 
of cold Galactic dust (T$\sim$15~K). 
We can note that the residual flux as well as the optical depth
become nearly constant for N$_{HI}\ge1.2$~$10^{21}$~H~cm$^{-2}$ since the 
contribution of the cold emission does not increase anymore.\\

In conclusion, in the diffuse medium, the spectrum of the dust associated with the
HI gas can be accuratly represented by a modified  
Planck curve with a $\nu^2$ emissivity law and a single temperature
(the residual emission is around 0, see Table \ref{tbl-1}) 
as it was already shown by Boulanger et al. 
(1996). For N$_{HI}$ larger than 5~$10^{20}$~H~cm$^{-2}$ ($A_{v}\ge$0.25), 
the cold dust (T$\sim$15~K) 
induces a significant submm excess with respect to the $\nu^2$ emissivity law.\\

\subsection{Dust emission spectrum in low HI column density regions}

In order to assess the reliability of the spectrum associated
with very diffuse regions, we explore now in detail, 
the variations of the HI spectrum in the low HI column density medium, 
where the contribution of the cold dust to the total emission is negligible
(N$_{HI}$$\le$4.5~$10^{20}$~H~cm$^{-2}$ i.e W$_{HI}$$\le$250~K~km~s$^{-1}$, see Table 1). 
This threshold is the same as in Boulanger et al. (1996).
We construct the spectrum of dust associated with
HI gas following Eq. \ref{eq_spec}, for various couples (l, k) 
of smaller HI sets of pixels (Table \ref{hi_bin}). \\

Fig. \ref{var_hi}a shows that the spectral shape of the four spectra is relatively constant
but the absolute level varies by a factor $\sim$1.6 between the two extremes.
The two spectra obtained in regions where $N_{HI}<2.5$~10$^{20}$~H~cm$^{-2}$ 
($W_{HI}<140$~K~km~s$^{-1}$) show
a significant variation of about 30$\%$. This variation cannot be explained by
the contribution of the cold dust since (1) it is not detected in these very
low HI column density regions (Table \ref{hi_bin})
and (2) the residual emission ($\nu~R_{\nu}$, see Sect. \ref{sect_HI_var} for the definition) 
is around zero. Moreover, cold dust would produce an excess at long wavelength
rather than a constant variation on the whole spectra.
In the next section, we show some evidence that these variations of the diffuse
medium spectra can be due to dust associated with the ionised gas, more
specifically from small scale structures in the WIM uncorrelated
with the distribution of HI gas. \\

\begin{table}[here]
\begin{center}
\caption{HI bins for the set of pixels l and k}
\label{hi_bin}
\begin{tabular}{|l|l|} \hline
W$_{HI}$ (l) & W$_{HI}$ (k) \\
K km s$^{-1}$ & K km s$^{-1}$ \\ \hline
$[0, 50]$ & $[50, 250]$ \\
$[0, 90]$ & $[90, 140]$ \\
$[0, 50]$ & $[50, 140]$ \\
$[0, 140]$ & $[140, 250]$ \\ \hline
\end{tabular}\\
\end{center}
\end{table}

\section{\label{sect_effect_hii} Effects induced by the diffuse ionized gas on the Far-IR/HI correlation }
To simulate the potential effect of the WIM Far-IR emission
on the Far-IR/HI correlation, we have first to evaluate the hydrogen column density
of the WIM. The only direct determination of the WIM
column density comes from dispersion measures in the direction of
high altitude pulsars, most of which located at $|b|<10\degr$ 
(Reynolds, 1989). However, these measures, which
give $N_{H^+}/N_{HI}~\sim~30~\%$,
do not necessarily apply to the local ISM. A more direct measurement
of the WIM at high Galactic latitude comes from Jahoda et al. (1990). 
They have measured in Ursa Major the HI emission through the 21 cm
line and the $H_{\alpha}$ intensity. They have obtained
$N_{HI}=0.7$~10$^{20}$~H~cm$^{-2}$ and $H_{\alpha}=0.2$~R.
The emission measure may be converted into column density assuming a
certain electron temperature and density. For $T_{e}=8000$~K
and $n_{e}=~0.08$~cm$^{-3}$ (Reynolds, 1991), 
we find for Ursa Major the same $N_{H^+}/N_{HI}$ ratio ($\sim30\%$)
as with dispersion measures.
In the following, we thus consider that this ratio do applied to the
high latitude sky.
These 30$\%$ contain the
correlated ($Nc_{H^+}$) and the non-correlated ($Nnc_{H^+}$)
H$^{+}$ components (by correlated, we mean spatially correlated with the HI gas).\\

The total hydrogen column density is defined as:\\
$N_H~=~N_{HI}~+~Nc_{H^+}~+~Nnc_{H^+}$.\\
We neglect the contribution of the molecular hydrogen since 
(1) it represents less
than 1$\%$ of the emission in regions where $W_{HI}<250$~K~km~s$^{-1}$
i.e. $N_{HI}<$4.5~10$^{20}$~H~cm$^{-2}$
(Tables \ref{tbl-1} and
\ref{hi_bin}) and (2) the fraction of H$_2$ 
gas is less than 1$\%$ in the very diffuse medium (Bohlin et al., 1978).
The correlated H$^{+}$ component can be written: $Nc_{H^+}=a~\times~N_{HI}$.
Since our knowledge on the non-correlated H$^{+}$ component is very poor, 
we choose to simulate it using two assumptions: \\
- (1) It is distributed in the sky like the 240~$\mic$
emission at the same longitude but opposite latitude. This assumption is 
arbitrary but ensures that the non-correlated WIM gas has the same scale 
properties and dependance with the Galactic latitude as the diffuse 
interstellar medium. We note that it is essential to use, for
$Nnc_{H^+}$, a map with small
scale structures to ensure the non-correlation with the HI component
at small scales. 
This map
is called \~{I$_{240}$}.\\
- (2) The non-correlated part of the WIM has an 
IR emissivity per hydrogen equal
to that of the HI gas, $\tau/N_{HI}=~10^{-25}~(\lambda/250\mic)^{-2}$ (Boulanger
et al., 1996). Computing $\tau$, defined as \~{I$_{240}$}$/B_{240}(T=17.5~K)$, 
gives the corresponding column density.\\

The map obtained with assumptions (1) and (2) 
is written \~{N$_{HI}$}. Thus, the uncorrelated H$^{+}$ component is represented
by: $Nnc_{H^+}=b\times$~\~{N$_{HI}$}. The constraint on $a$ and $b$ is: $a+b=30\%$. \\

\begin{figure*}
\begin{minipage}{7.7cm}
\epsfxsize=8.cm
\epsfysize=7.cm
\hspace{-0.5cm}
\vspace{0.4cm}
\epsfbox{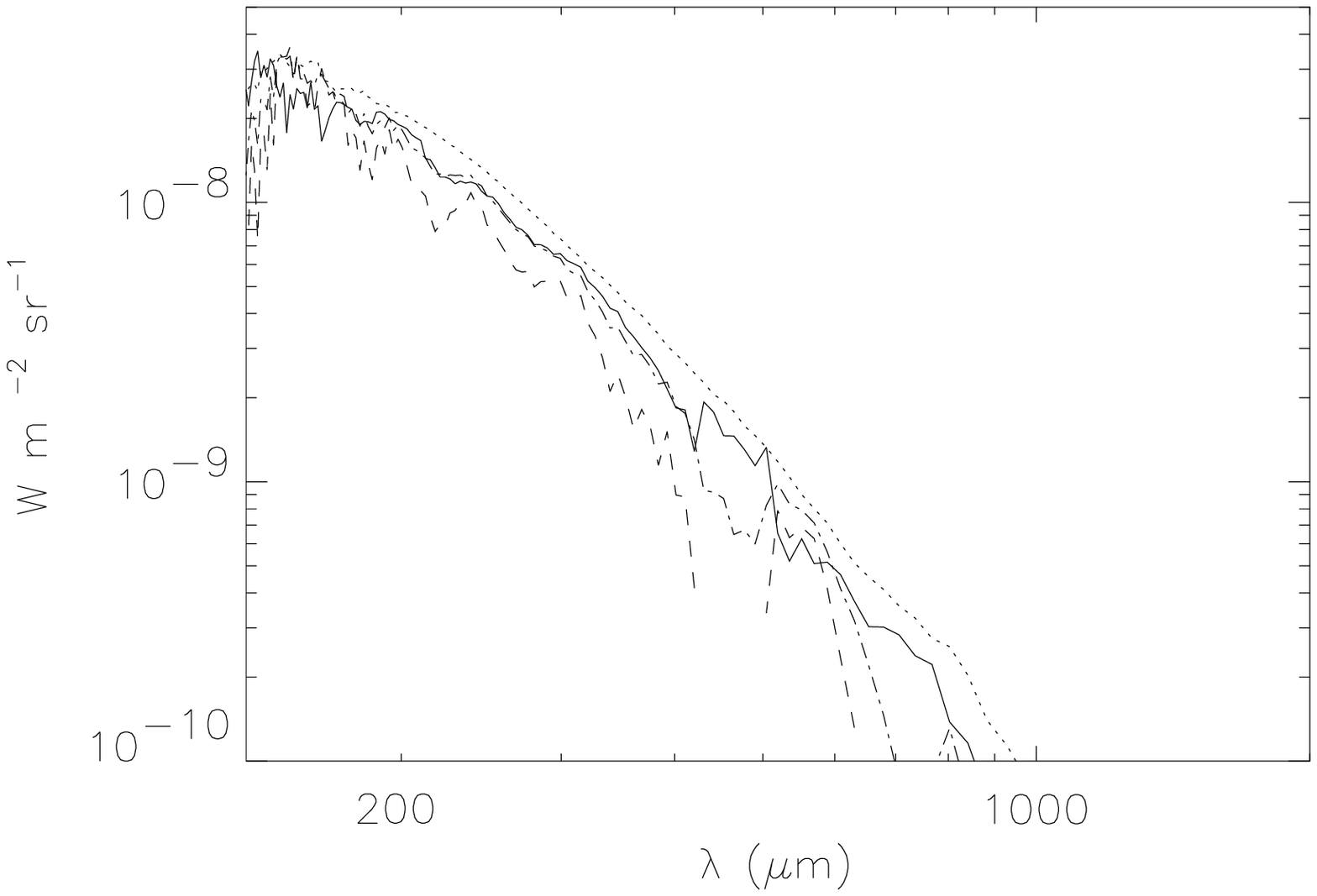}
\end{minipage}
\begin{minipage}{7.7cm}
\epsfxsize=8.cm
\epsfysize=7.cm
\hspace{-0.5cm}
\vspace{0.4cm}
\epsfbox{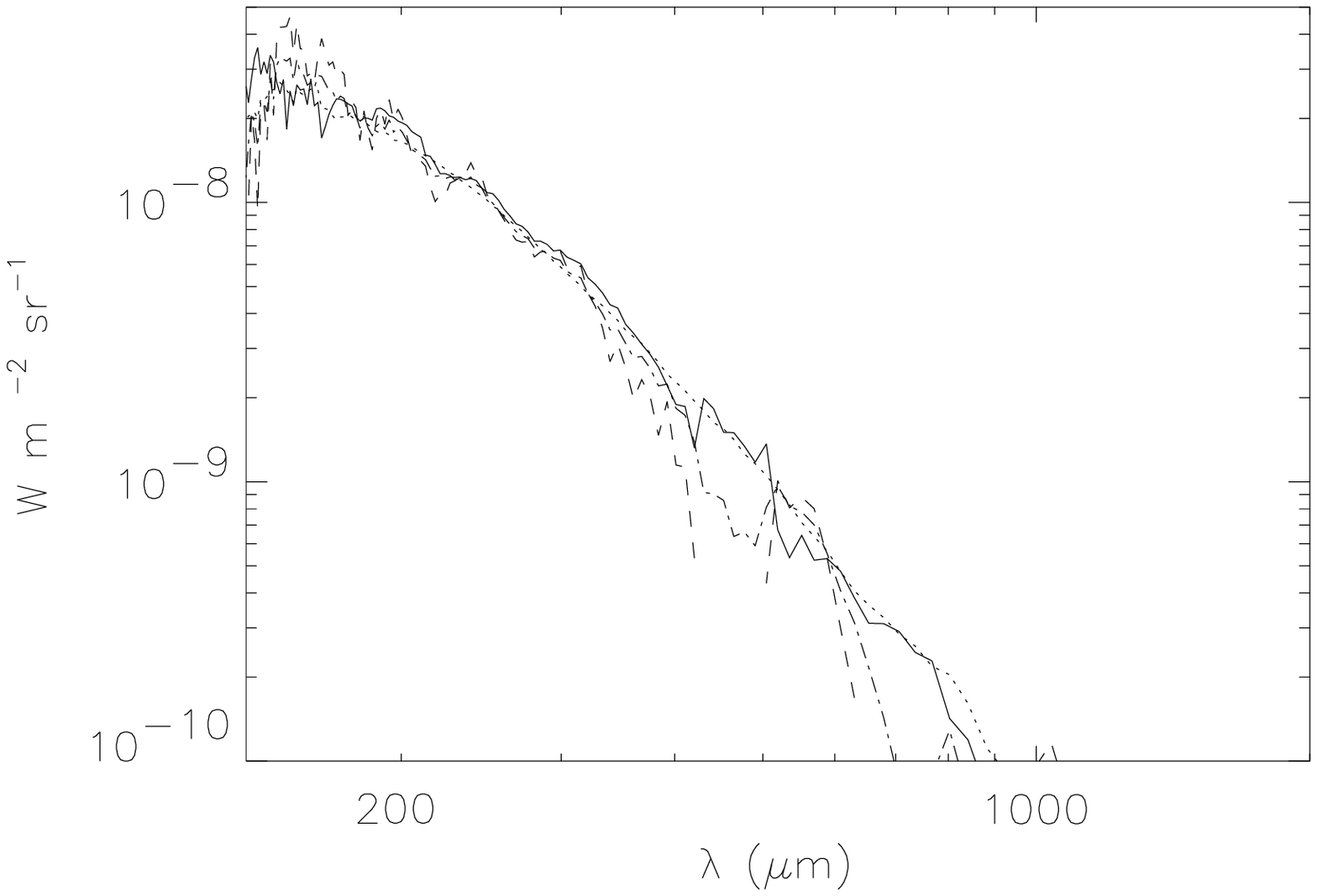}
\end{minipage}\\
\vspace{-1.cm}
\caption{\label{var_hi}  (a) Spectra of dust associated with HI gas 
(normalised at N$_{HI}$=10~$^{20}$~H~cm$^{-2}$) computed following 
the Eq. \ref{eq_spec} for different sets of pixels (Table \ref{hi_bin}). 
(b) Spectra of the dust computed in the same way as (a) and for the same sets. 
The normalisation
is no more by HI atom but by the total hydrogen column density defined as 
$N_H~=~N_{HI}~+~Nc_{H^+}~+~Nnc_{H^+}$ with 
$Nnc_{H^+}=b~\times$~\~{N$_{HI}$}, $Nc_{H^+}=a~\times N_{HI}$,
$a=0.09$ and $b=0.21$ (see Sect. \ref{sect_effect_hii}).
The dotted line corresponds to the $W_{HI}$ sets of pixels [0, 140] and [140, 250], 
the continuous line
to [0, 90] and [90, 140], the dashed line to [0, 50] and [50, 140] 
and the dashed-dotted line to [0, 50] and [50, 250]~K~km~s$^{-1}$.
Spectra have been truncated below 135~$\mic$ because of the poor signal to noise ratio.
They have been smoothed to a resolution of 4~cm$^{-1}$.}
\end{figure*}

Dust spectra are computed following Eq. \ref{eq_spec}, using 
$<N_{H}>_{l,k}$ rather than $<N_{HI}>_{l,k}$,
for the same HI sets as in the previous section (Table \ref{hi_bin})
and for different values of $a$ and $b$ 
with a step equal to 1$\%$ and the constraint $a+b=30\%$. 
We quantify the dispersion 
between the spectra by computing the total rms of the difference between the four spectra
and the mean spectrum, in the wavelength range [200, 300] $\mic$. 
Fig. \ref{var_hi}b shows the
result for the minimal dispersion, obtained with b$\sim$21$\%$. We see no more
differences between the absolute levels of the spectra. 
They are all stabilized on the mean spectrum.
The same conclusion is reached with other HI
sets of pixel as for example: 
l=[0, 50] and k=[50, 190], l=[0, 190] and k=[190, 250], l=[0, 70] and 
k=[70, 160], l=[0, 100] and k=[100, 180]~K~km~s$^{-1}$.
This simulation is only illustrative but we think that it has a quantitative
significance provided that the spatial distribution of the uncorrelated WIM 
shares the same morphological properties (ratio between small scale structures and large
scale Galactic gradient) than our ``DIRBE'' template. This is
demonstrated by the fact that the optimal $b$ value is very
stable when we rotate the $H_{nc}^+$ template with the
Galactic longitude as long as this template stays uncorrelated with the HI
component at small scales. 
This simulation leads to the following
proportion of uncorrelated and correlated H$^+$ gas of  
70$\%$ and 30$\%$ which are in agreement with the Reynolds et al. (1995) determination
in a small region.\\

The emissivity of the spectrum of dust associated with the HI gas 
(including the $H^+_c$ component) 
that we deduce is:
\begin{equation}
\label{emiss_HI}
 \tau/N_{HI}= 8.7\pm 0.9 \quad 10^{-26} (\lambda/250\mic)^{-2} \quad cm^2
\end{equation}
with a temperature of 17.5 K. 

This emissivity value is compatible with the one derived in
Boulanger et al. (1996).\\

\section{\label{sect_hii_spec} Emission spectrum of the diffuse H$^{+}$ component}

\subsection{Detection}

The IR emission from any dust associated with the WIM would follow a $csc(b)$ variation, 
like that from any diffuse component in the Galactic disk. 
Based on such an hypothesis, Boulanger et al. (1996) have found in the FIRAS data
a residual Galactic emission (after the removal of the HI component) consistent with an
emission spectrum like that of the HI gas for $N_{HI}\sim$~4~10$^{19}$~H~cm$^{-2}$.
They consider this as an upper limit
for the contribution of the H$^{+}$ component since they could have measured emission
from low contrasted molecular clouds. \\

To search for any residual diffuse component after the removal of the HI contribution,
we use all sky FIRAS maps at $|b|>20\degr$. We keep only 
pixels for which $|\Delta S|<1~\sigma$ (see Sect. \ref{sect_data_prep}).
This criterium is over restrictive (we keep only 25.6$\%$ of the sky) but ensure that 
the remaining pixels are mostly coming from the diffuse medium. \\


For each selected FIRAS spectrum $F_{\nu}(i,j)$, 
we remove the HI related emission 
following the equation:
\begin{equation}
\Delta F_{\nu}(i,j) = F_{\nu}(i,j) - N_{HI}(i,j) \times I_{HI}(\nu)
\label{eq_resid_firas}
\end{equation}
where N$_{HI}$ the HI column density and 
$I_{HI}(\nu)$ is computed from
$\nu I_{HI}(\nu)=~\tau/N_{HI}~\times~B_{\nu}(17.5 K)$
where $\tau/N_{HI}$ is the HI dust emissivity given Eq. \ref{emiss_HI}
and $ B_{\nu}(17.5K)$ the Planck curve at 17.5~K in W~m$^{-2}$~sr$^{-1}$. \\

\begin{figure}[top]  
\epsfxsize=8.cm
\epsfysize=7.cm
\hspace{2.cm}
\vspace{0.0cm}
\epsfbox{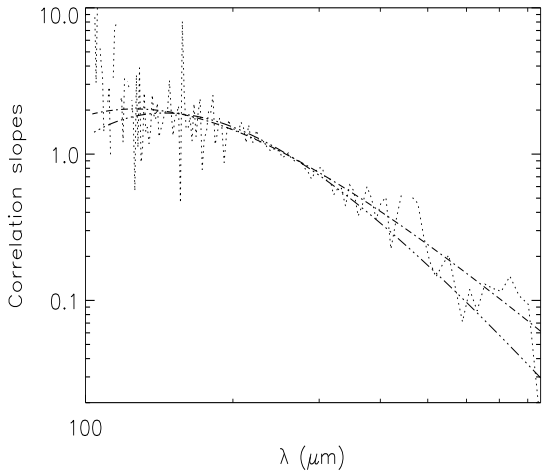}
\caption{\label{hii_spec} 
Correlations slopes
of the residual Galactic emission (dotted line)
after the main HI component has been removed
(slopes have been computed using the
residual Galactic emission in $I_{\nu}$).
The dash-dotted line
represents a modified Planck curve with $\alpha$=1, T=29.1~K and 
the dot-dot-dot-dashed line a modified Planck curve with $\alpha$=2 and T=20~K.
This spectrum is the first detection
of the WIM dust emission}
\end{figure}

We first search for any residual emission by correlating the Galactic latitude emission 
profiles of $\Delta F_{\nu}$ at each FIRAS wavelength,
with the mean latitude emission profile computed in the wavelength range [200-350] $\mic$.
This correlation method is probably more accurate because it avoids the 
problems with the cosecant method linked to the local bubble.
The spectrum obtained with these correlations 
is presented on Fig. \ref{hii_spec}. 
This emission is clearly non zero and is detected at a 10$\sigma$ level
in the [200-350]~$\mic$ band.
The spectrum can be fitted by a modified Planck curve with a $\nu^\alpha$ emissivity law.
The best fit is obtained for $\alpha$=1 and T=29.1~K.
The value $\alpha$=2 is also compatible with the data. In that case, we have
T=20.0~K, which is significantly higher than the 
temperature of dust associated with HI gas.\\

To check the Galactic origin of this residual emission,
we fit the latitude profiles computed over the selected
regions and for each FIRAS wavelength with cosecant variations:
\begin{equation}
\Delta F_{\nu}(|b|) \propto D_{\nu}cosec|b|
\label{eq_cosec}
\end{equation}
The slopes $D_{\nu}$ measured at each wavelength form
a spectrum which has the same temperature and shape 
as the spectrum obtained with the correlation (Fig. \ref{hii_spec}). 
Therefore, we conclude
that the spectrum derived with the correlation has a Galactic origin. \\

The ISM is known to contain structures over a very wide range of scales with inhomogeneous
physical conditions. Within the large FIRAS beam (7$\degr$), 
small scale molecular clouds or HII regions can be diluted 
and thus not detectable using our criteria. In order  
to check whether such small scale structures contaminate the residual spectrum 
of Fig. \ref{hii_spec}, we use the same two methods (correlation and cosecant 
variations) to derive the spectrum  using different cuts
for the selection of the pixels
(either more or less restrictive than above). 
The spectra obtained with correlations or cosecant laws are very stable. 
This supports the idea that the residual Galactic emission cannot
be due to low contrasted molecular clouds or HII regions. \\
Moreover, to check whether this residual Galactic emission 
is not dominated by a particular region of the sky, we repeat the same cosecant fit 
analysis
in 4 bins of longitude of roughly equal size (6$\%$ of the sky each). The small number
of pixels does not allow to have a better sampling. We see, 
in Table \ref{tbl-var-hii-long}, that the residual Galactic 
component is present in the 4 different bins with small variations. 
Therefore, we can exclude
this residual Galactic component to come from a particular region of the sky.\\

\begin{table}
\begin{center}
\caption{\label{tbl-var-hii-long}
Cosecant slopes ($\nu~D_{\nu}$)
of the residual Galactic emission
averaged in the wavelength band [200,350]~$\mic$ and
computed in different Galactic longitude areas.}
\begin{tabular}{|c|c|} \hline
Galactic longitude & Residual Galactic emission \\
range (in $\degr$) & (10$^{9}$ W m$^{-2}$ sr$^{-1}$ )  \\ \hline
0 - 85 & 9.4$\pm$0.5 \\
85 - 170 & 6.4$\pm$0.4 \\
170 - 235 & 9.9$\pm$0.6 \\
235 - 360 & 10.7$\pm$0.7 \\\hline
\end{tabular}\\
\end{center} 
\end{table}

The residual Galactic spectrum corresponds to material which is 
uncorrelated with the HI gas (by construction). 
However, it is Galactic because it follows a cosecant law.
We propose that this is the first detection of the submm emission 
of dust associated with the WIM. 

\subsection{Normalisation and emissivity}
To normalise the WIM far-IR spectrum, we make use of the cosecant law measured
for $H_{\alpha}$ emission at high Galactic latitude.
The slope of the H$_{\alpha}$
cosecant law is egal to 1 R (Reynolds et al., 1984). 
This gives (following the results obtained
in Sect. 4: 70$\%$ of 1R) 0.7 R for the non-correlated
$H^+$ gas. Using a conversion factor 
$I_{H \alpha}~(R)~=~1.15~\times~N_H^+/10^{20}$~cm$^{-2}$
(derived from Osterbrock, 1989),
we obtain $Nnc_{H^+}=~6.1$~$10^{19}$~cm$^{-2}$. This column density is
close to that derive from the pulsar observations (Reynolds, 1989). \\
Using this normalisation, the emissivity per H$^{+}$ ions is: 
$\tau/N_H^+=~3.8\pm0.8$~$10^{-26}~(\lambda/250\mic)^{-1}$~cm$^2$.
The uncertainty corresponds to the 21 $\%$
dispersion observed in the different longitude bins  
(Table \ref{tbl-var-hii-long}). With a $\nu^2$
emissivity law, we find $\tau/N_{H^+}=~1.0\pm 0.2$~$10^{-25}~(\lambda/250\mic)^{-2}$
cm$^2$. This emissivity value agrees within error bars with the
emissivity of dust associated with HI gas. \\

We have not taken into account until now of the uncertainty on the
HI dust emission spectrum.
The removal of the HI maximal spectrum,
defined by $\tau/N_{H^+}=~9.6~$~$10^{-25}~(\lambda/250\mic)^{-2}$ (see Eq. \ref{emiss_HI})
and T=17.5 K, only decreases the emissivity of the WIM dust component 
by 30$\%$; the removal of the minimal spectrum, defined by 
$\tau/N_{H^+}=~7.8$~$10^{-26}~(\lambda/250\mic)^{-2}$
and T=17.5 K, increases the emissivity by 24$\%$.

The spectrum derived from the slopes of the cosecant law 
(which measures the Galactic emission in a direction perpendicular
to the Galactic disk over a significantly large area)
corresponds to the submm dust emission of an HI equivalent 
column density of 10$^{20}$~cm$^{-2}$ (this value is estimated using $D_{\nu}$ and 
$I_{HI}(\nu)$),
which is higher than the tentative detection of Boulanger et al. (1996) 
due to the difference in the
HI dust spectrum used for the normalisation. Since 
this column density of 10$^{20}$~cm$^{-2}$ is roughly the Reynolds's estimate
of the total H$^+$ column density, our result is consistent with a dust 
abundance in the WIM which does not differ much from
that of the HI gas. We reach the same conclusion using the minimal
or maximal HI spectrum.
However, this conclusion is still preliminary since the Reynolds's 
estimate is made in parts of the sky which are different from those used here.
Future comparison of the Far-IR dust emission with the Wisconsin H-Alpha
Mapper survey (WHAM, Tufte et al., 1996; Reynolds et al., 1998) will allow to give a 
more precise estimate on the dust abundance in the WIM. \\

\subsection{First interpretation}

\begin{figure}[top]  
\epsfxsize=8.cm
\epsfysize=7.cm
\hspace{2.cm}
\vspace{0.0cm}
\epsfbox{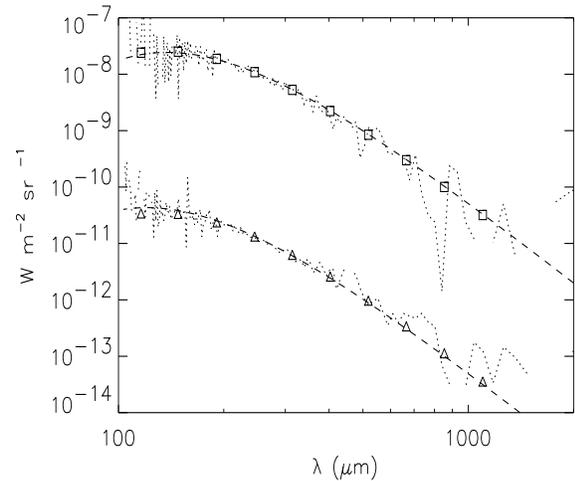}\\
\vspace{-0.5cm}
\caption{\label{hii_hi_models}Upward curves: HI dust emission spectrum (dotted line), 
HI synthetic spectrum 
(corresponding to the emissivity and temperature given in Sect. \ref{sect_effect_hii}, 
dashed line) and HI emission 
model (square) for $N_{HI}~=~10^{20}$~cm$^{-2}$.
Downward curves: H$^{+}$ dust emission spectrum (dotted line, divided
by 1.5~$10^{3}$),
H$^{+}$ synthetic spectrum (corresponding to the emissivity and temperature given in Sect. 
\ref{sect_hii_spec}, dashed line) and H$^{+}$ emission model (triangle) for
$N_{H^+}~=~10^{20}$~cm$^{-2}$.}
\end{figure}

It is extremely difficult to disentangle whether the WIM spectrum
follows a $\nu^2$ or a $\nu^1$ emissivity law because of the poor signal to noise ratio
of our spectrum at long wavelength (Fig. \ref{hii_spec}). 
However, this spectrum has a definitively higher temperature
than the spectrum of dust associated with HI gas. \\
Dust grains in the WIM are expected 
to have smaller sizes due to grain shattering in 
grain-grain collisions (Jones et al., 1996, Fig. 17).
We have used the dust model developped by D\'esert et al. (1990) to test the effect of the 
size distribution on the Far-IR spectrum. First, a standard Big Grains composition and size
distribution (abundance in mass m/m$_{H}$=~5~10$^{-3}$ {\footnote{This abundance in mass
is lower than the one given in Desert et al., 1990 due to the difference of
calibration between the IRAS and DIRBE/FIRAS instruments}},
silicates with dark refractory mantle composition, distribution in size following 
a 2.9 power law with a$_{min}$=15~nm and a$_{max}$=110~nm, and a density of 3~g/cm${^3}$)
remarquably reproduce the dust emission spectrum associated with the HI gas (Fig. \ref{hii_hi_models}, upward curve).
A smaller cutoff in the size distribution (with a$_{max}$=30~nm, 
which is compatible with what is predicted
in Jones et al., 1996) is needed to reproduce the spectrum of the 
WIM (Fig. \ref{hii_hi_models}, downward curve).
This comparative analysis suggests that the difference of temperature between the HI and H$^{+}$ 
diffuse media can be explained by the erosion of the dust grains due to the collisions.

\section{The extragalactic component}
\subsection{\label{sect_CFIRB_LH} FIRAS spectra in low HI column 
density regions: evidence for an extragalactic component}

\begin{figure}  
\epsfxsize=8.cm
\epsfysize=7.cm
\hspace{2.cm}
\vspace{0.0cm}
\epsfbox{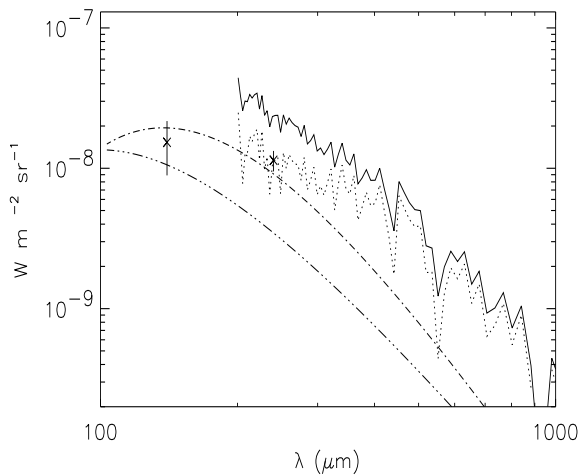}\\
\caption{\label{firback_lh_fir} Mean FIRAS (continuous line),
HI (dash-dot) and H$^{+}$ (dash-dot-dot-dot) spectra
in regions where N$_{HI}<$9.1~10$^{19}$~H~cm$^{-2}$. The residual emission (dot line), 
obtained by removing
from the mean FIRAS spectrum the HI and H$^{+}$ components, dominates the FIRAS emission at high
latitude in low HI column density regions. The typical uncertainties for this residual emission
are around 2.6 10$^{-9}$~W~m$^{-2}$~sr$^{-1}$ for $200<\lambda<400~\mic$, 
1.75~10$^{-9}$~W~m$^{-2}$~sr$^{-1}$ for $400<\lambda<600~\mic$, and
2.5~10$^{-10}$~W~m$^{-2}$~sr$^{-1}$ for $\lambda>600 \mic$.
Also reported are our determinations of the DIRBE CFIRB at 140 and 240~$\mic$ (cross points, see
Table \ref{CFIRB_DIRBE}).}
\end{figure}

Puget et al. (1996) have found in the residual emission after 
the removal of the HI correlated
emission, an isotropic component which could be the Cosmic Far Infrared Background (CFIRB)
due to distant Galaxies. At $|b|>40\degr$, at least one third of the emission at 500 $\mic$
comes from this isotropic component. Thus, at high Galactic 
latitude in very low HI column density regions,
this component should dominate the FIRAS emission.
\\
We have selected pixels for which N$_{HI}<$9.1~10$^{19}$~H~cm$^{-2}$ and b$>$40$\degr$
(33 pixels). 
These pixels are located near and in the Lockman Hole and represent 0.54$\%$ of the sky. 
For this set of pixels, the mean HI column density is equal to 8 10$^{19}$ H cm$^{-2}$.
In this region, the non-correlated H$^{+}$ column density is estimated using 
$N_{H^+}$/~N$_{HI}$=30$\%$ (Jahoda et al., 1990)
which gives  $Nnc_{H^+}$/~$N_{HI}$=21$\%$ (see Sect. \ref{sect_effect_hii}).
We clearly see (Fig. \ref{firback_lh_fir}) that the emission of dust associated with the 
HI and H$^{+}$ components
cannot account for the FIRAS spectrum, especially at $\lambda$$>$300 $\mic$. 
The presence of another component is evident.  
The level of this component is the same as the residual FIRAS emission 
obtained in Puget et al. (1996) before they subtract the estimated H$^{+}$ 
and is in very good agreement with the Fixsen et al. (1998) determinations.
Puget et al. (1996) have shown that the most likely interpretation for this component is 
the Cosmic Far-Infrared Background (CFIRB) due to the integrated light of distant galaxies.
Despite the fact that it is not physical to consider a unique temperature for this background, 
it can be analytically represented in the same way as in Fixsen et al. (1998).
The best fit, valid between 200 and 2000~$\mic$, 
\begin{equation}
\label{analy_CFIRB}
I(\nu)=8.80 \times 10^{-5} (\nu / \nu_0)^{1.4} B_{\nu}(13.6 K)
\end{equation}
where $\nu_0$=100~cm$^{-1}$, is presented on Fig. \ref{51_CFIRB}, together
with the range of determination of Fixsen et al. (1998). 
The uncertainties on the fit will be discussed
in a forthcoming paper (Gispert et al., 1999).\\

\begin{table*}
\begin{center}
\caption{\label{CFIRB_DIRBE}
Contribution of the different components to the DIRBE 140 and 240~$\mic$ emissions in 
the Lockman Hole region where the mean HI column density is 8~10$^{19}$~H~cm$^{-2}$.
Res2 corresponds to the CFIRB values.}
\begin{tabular}{|l|c|c|} \hline
Component & 140$\mic$ & 240$\mic$ \\ 
& 10$^{8}$ W m$^{-2}$ sr$^{-1}$ & 10$^{8}$ W m$^{-2}$ sr$^{-1}$ \\ \hline
Total &4.53$\pm$0.57  & 2.40$\pm$0.15 \\
HI+H$_c^+$ & 1.94$\pm$0.19 & 0.91$\pm$0.09 \\ 
Res1=Total - (HI + H$_c^+$) & 2.59$\pm$0.60 & 1.49$\pm$0.17 \\ \hline
H$_{nc}^+$ & 1.06$\pm$0.22 & 0.35$\pm$0.07 \\
Res2=Total - (HI + H$_c^+$) - H$_{nc}^+$ & 1.53$\pm$0.64 & 1.14$\pm$0.19 \\ \hline
Residual $^{1}$& 2.50$\pm$0.70 & 1.36$\pm$0.25 \\ 
Residual $^{2}$& 3.19$\pm$0.43 & 1.67$\pm$0.17 \\ \hline
\end{tabular}\\
\end{center}
\hspace{3.8cm} $^{1}$ Hauser et al., 1998
\par\medskip
\hspace{3.8cm} $^{2}$ Schlegel et al., 1998\\
\end{table*} 

The CFIRB can also be determined at 240 and 140~$\mic$ using DIRBE data for the same part
of the sky. Results are presented in Table \ref{CFIRB_DIRBE} together with previous 
determinations. In this Table, 
uncertainties on total emissions are the dispersions measured for the
selected pixels, 
uncertainties on the HI dust emissions are given by Eq. \ref{emiss_HI}, and
uncertainties on the WIM dust emissions correspond to
the 21$\%$ dispersion
observed in the four bins of Galactic longitude (Table \ref{tbl-var-hii-long}).
Our residual emissions (Res2 in Table \ref{CFIRB_DIRBE}) are significantly
lower than the ones reported in Hauser et al. (1998) and
Schlegel et al. (1997) since they have neglected the contribution of the dust
associated with the WIM. We clearly see that it is essential
to take into account the dust emission associated to the WIM below 240~$\mic$.
In FIRAS data at 240~$\mic$, the CFIRB emission (Fig. \ref{firback_lh_fir}) 
is very close to our  
DIRBE value. At 140~$\mic$, the comparison between DIRBE and FIRAS data is not 
possible due to the considerable increase of the FIRAS data noise (that is why 
we have prefered to cut the spectrum at 200~$\mic$ in Fig. \ref{firback_lh_fir}).

\subsection{\label{sect_CFIRB_iso} Isotropy}
We test in this part the isotropy of the CFIRB on large scales 
since we do not know the spatial distribution of the WIM dust
emission at small angular scales. This test at large scales is important to 
detect potential systematic effects caused by an inacurate subtraction of
ISM dust emissions. As expected from our detection of the WIM
dust emission, we show that the CFIRB is isotropic only
if we consider the emission of dust associated with the H$^+$ component.\\

\begin{figure}  
\epsfxsize=8.cm
\epsfysize=7.cm
\hspace{2.cm}
\vspace{0.0cm}
\epsfbox{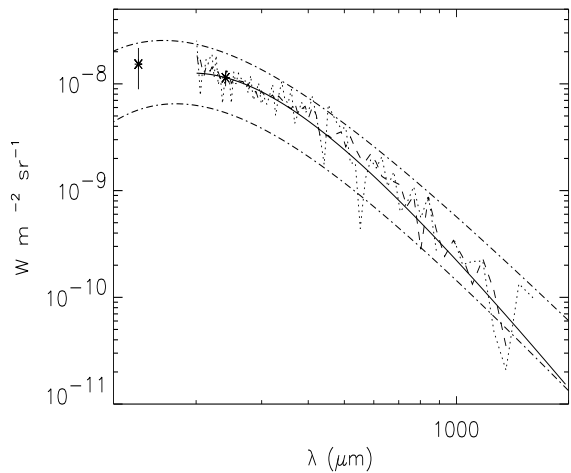}\\
\caption{\label{51_CFIRB} CFIRB emission computed on 51$\%$ (dashed line)
and on 0.54$\%$ of the sky (dotted line). The 
continuous and dotted-dashed lines correspond to the analytic 
forms of the CFIRB given by Eq. \ref{analy_CFIRB} and
Fixsen et al. (1998) respectively. Cross points correspond to our DIRBE CFIRB
values at 140 and 240~$\mic$ (Res2 of Table \ref{CFIRB_DIRBE}).}
\end{figure}

Thus, we have to compute the CFIRB spectrum on a 
fraction of the sky as large as possible. For that, we prefer to use a Galactic template
based on far-IR emissions rather than HI gas, (1) to take into account
variations in the dust temperature from place to place and (2) to avoid the large hole
in the southern hemisphere of the Leiden/Dwingeloo survey.
We combine the DIRBE 140 and 240~$\mic$ 
data with FIRAS spectra. First, we extract from DIRBE data (at 7$\degr$ resolution) 
a Galactic emission template. Then, 
the DIRBE Galactic emission is extrapolated to longer wavelengths and subtracted 
from each individual selected FIRAS spectrum to derive the CFIRB emission and
test its isotropy.\\

The two (140 and 240~$\mic$) Galactic DIRBE templates, $I_G$,  
are computed by removing the CFIRB (Table \ref{CFIRB_DIRBE}) from the DIRBE emissions. 
Assuming a $\nu^2$ emissivity law, temperature
and optical depth of each pixel are estimated using the two DIRBE Galactic templates. 
The residual FIRAS emission is computed in the following way. 
First, we discard pixels located in known
molecular clouds or HII regions. We keep pixels with
$|~\Delta S|<3~\sigma$ and work at $|b|>15 \degr$ (51$\%$ of the sky).
Then, for each selected pixel with its associated temperature,  
we derive the ratio $R_{\nu}$ between the modified Planck curve computed at each FIRAS 
wavelength
and the modified Planck 
curve computed at 240~$\mic$. Finally, the residual emission is computed
in the following way:
\begin{equation}
Res_{\nu}(i,j)= F_{\nu}(i,j) - R_{\nu} \times I_G(i,j)
\end{equation}
The mean residual emission spectrum is shown
Fig. \ref{51_CFIRB} together with the spectrum obtained on 0.54$\%$ of the sky 
and its analytical
determination (Eq. \ref{analy_CFIRB}). The two spectra agree very well.

\begin{figure}  
\epsfxsize=8.cm
\epsfysize=7.cm
\hspace{2.cm}
\vspace{0.0cm}
\epsfbox{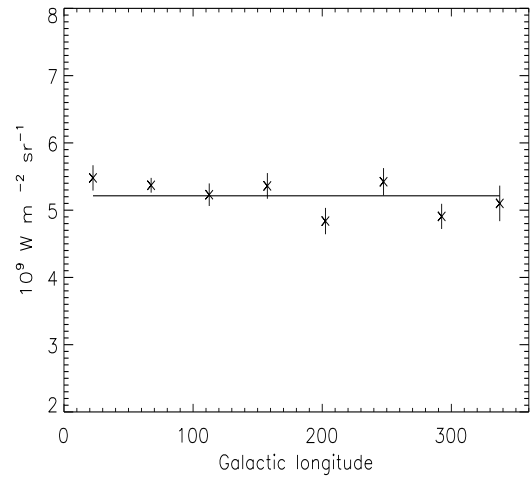}\\
\caption{\label{CFIRB_long} Variation of the FIRAS residual emission 
averaged in the [300, 609]~$\mic$ band
with the Galactic longitude. 
The error bars are statistical errors (1$\sigma$).}
\end{figure}

\begin{figure*}
\begin{minipage}{7.7cm}
\epsfxsize=8.cm
\epsfysize=7.cm
\hspace{-0.5cm}
\vspace{0.4cm}
\epsfbox{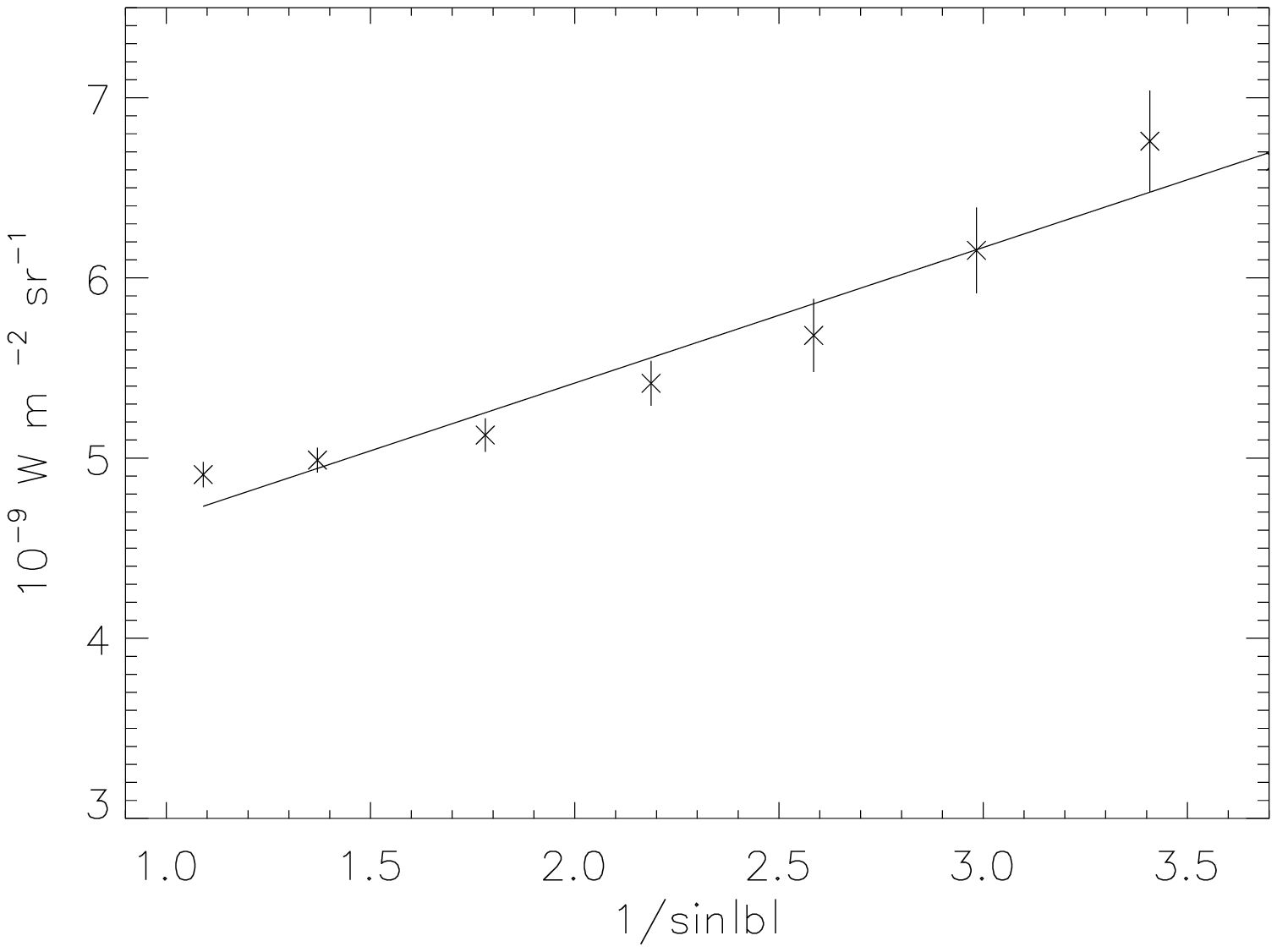}
\end{minipage}
\begin{minipage}{7.7cm}
\epsfxsize=8.cm
\epsfysize=7.cm
\hspace{-0.5cm}
\vspace{0.4cm}
\epsfbox{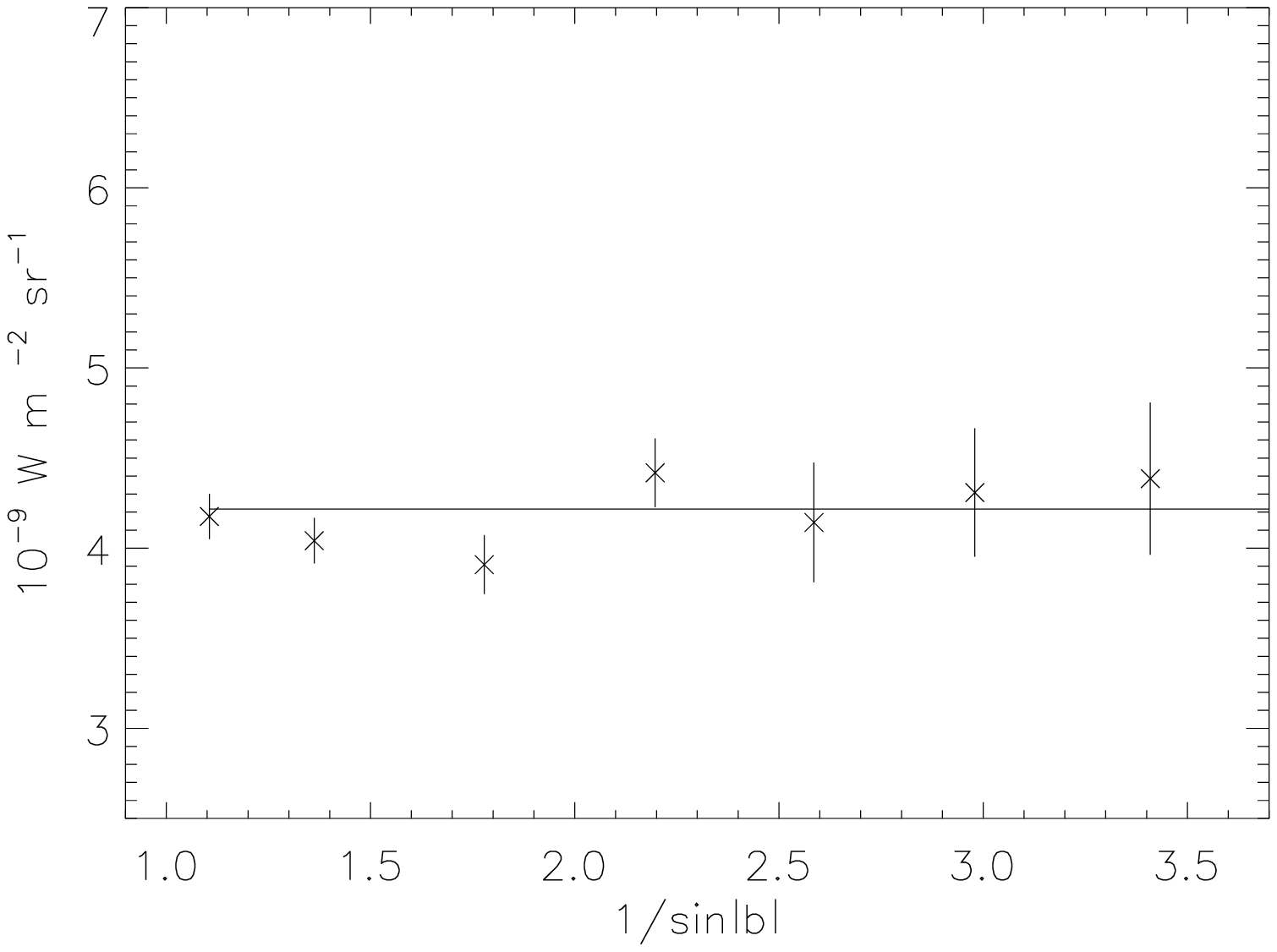}
\end{minipage}\\
\vspace{-1.cm}
\caption{\label{CFIRB_lat} a. Variation of the FIRAS residual emission 
averaged in the [300, 609]~$\mic$ band with the Galactic latitude.
b. same as (a) but the FIRAS residual has been computed by taking into account dust 
emission from the WIM.}
\end{figure*}

The isotropy of our residual emission is addressed 
by looking at its variation with the Galactic longitude or latitude.
The variation of our residual with the Galactic longitude is derived by  computing
the mean residual emission (in the [300, 609]~$\mic$ band) in 8 different longitude bins 
(each bin representing around 3.5$\%$ of the sky). 
The profile, shown in Fig. \ref{CFIRB_long}, does not show any particular Galactic structure. 
To test the isotropy with the Galactic latitude, we compute the emission profile
versus the latitude (cosecant variation) of
our residual emission (Fig. \ref{CFIRB_lat}a). We clearly see
a residual Galactic component. The slope of the fitted cosecant law is 
$\sim$~7~10$^{-10}$~W~m$^{-2}$~sr$^{-1}$ 
which is 4 times smaller than the slope of the H$^+_{nc}$ cosecant component.\\

The pixels selected before represent only the diffuse medium but contain both the
neutral and ionised emission and thus cannot be very well represented by a single
temperature component. Therefore, our residual FIRAS emission could be affected by 
an inaccurate subtraction of the interstellar medium emissions.
To test the effect of the presence of the WIM dust emission on our
residual emission, we apply the same method as previously but remove from the
DIRBE data, before determining the temperatures, the H$_{nc}^+$ dust emission defined
as $Enc_{H^+}(i, j, \nu) = b \times$ \~{N$_{HI}$}$(i, j)
\times Inc_{H^+}(\nu)$ with $b=0.21$ (see Sect. \ref{sect_effect_hii}). $Inc_{H^+}(\nu)$
represents the non-correlated $H_{nc}^{+}$ spectrum derived in Sect. \ref{sect_hii_spec}.
The FIRAS residual emission is computed in the following way:
\begin{equation}
Res_{\nu}(i,j)= F_{\nu}(i,j) - R'_{\nu} \times I_G'(i,j) - Enc_{H^+}(i,j, \nu)
\end{equation}
where $R'_{\nu}$ and 
$I_G'$ represent the new ratio and DIRBE 240~$\mic$ neutral Galactic 
template respectively, obtained after removing the contribution of the dust emission
associated with the non-correlated HII component.
The mean residual emission is in very good agreement with the previous estimates
(Fig. \ref{51_CFIRB}). Moreover,
we see no more residual cosecant variations (Fig. \ref{CFIRB_lat}b):
the residual component of Fig. \ref{CFIRB_lat}a was due to the non-correlated $H^{+}$
emission. \\

In conclusion, we 
clearly show that the main uncertainty to the CFIRB determination on individual pixels
of the sky comes from dust emission associated with the non-correlated H$^+$ component.

\section{\label{sect_cl} Conclusions}
A new analysis of the correlation between gas and dust at high Galactic
latitude has been presented. For the first time, we are able to present a 
decomposition of the Far-IR emission over the HI, H$^+$ and H$_2$
gas with the determination of the dust emission spectrum for each of
these components. This decomposition is important to study
the evolution of interstellar dust as well as to give
new constraints on the CFIRB (spectrum and test of isotropy).
 
The data used in this analysis are the COBE far-IR data and the Leiden-Dwingeloo HI survey.
Our decomposition is based on the IR/HI correlation and
DIRBE and FIRAS IR colors. 
The main quantitative results of this work follow:\\

1) Dust emission spectrum associated with the HI gas:\\
We confirm that the emission spectrum of dust in
HI gas is well fitted by a modified Planck curve with
a $\nu^2$ emissivity law and a temperature of 17.5~K. However,
the emissivity normalised per H atom varies 
by about 30$\%$ depending on the pieces of the sky
used to compute the spectrum.
We show that this variation may be explained by the
emission of the dust associated with the H$^+$ component.
Taking into account this contribution, we derive a new value of
the dust emissivity
nomalised per hydrogen atom:
$\tau/N_{HI}$=~8.7$\pm$~0.9~$10^{-26}~(\lambda/250\mic)^{-2}$~cm$^2$ with a temperature of 
17.5~K, slightly lower than the previous determination of
Boulanger et al. (1996).\\

2) Dust emission spectrum associated with the H$^+$ gas:\\
The existence of far-IR emission from the WIM is demonstrated
by the latitude dependance of the residuals of the IR/HI correlation.
This cosecant variation gives a spectrum which is clearly different 
from the HI dust spectrum. 
This component is detected at a 10$\sigma$ level in the [200-350]~$\mic$
band. Dust associated with the WIM has an emissivity 
$\tau/N_{H^{+}}=$~3.8$\pm$~0.8~$10^{-26}~(\lambda/250\mic)^{-1}$~cm$^2$ 
with a temperature of 29.1 K. 
With a spectral index equal to 2, the emissivity law becomes
$\tau/N_H^+=$~1.0~$\pm$~0.2~$10^{-25}~(\lambda/250\mic)^{-2}$~cm$^2$
with a temperature of 20~K.
The variation in dust spectrum from the HI to the WIM dust component
can be explained by only changing the upper cutoff
of the Big Grain size distribution from 0.1~$\mic$ to 30~nm.\\

3) Cosmic Far-IR Background (CFIRB):\\
FIRAS spectra in low HI column density regions (0.54$\%$ of the sky) clearly show
the presence of a component which is not associated with the HI and 
H$^{+}$ gas and which is interpreted, following Puget et al. (1996), as the 
CFIRB due to the integrated light of distant galaxies. 
The determination of this component on 51$\%$ of the 
sky confirms its isotropy at large scale
on a suitable fraction of the sky. 
With our determination of the WIM contribution to the far-IR sky emission, we
find for the CFIRB values at 140 and 240~$\mic$, $1.53\pm0.64$~10$^{-8}$ and
$1.14\pm0.19$~10$^{-8}$~W~m$^{-2}$~sr$^{-1}$ respectively, which are significantly
lower than the Hauser et al. (1998) determinations. The contribution of
the IR dust emission from the WIM relative to the CFIRB is negligible at
longer wavelengths.

\acknowledgements
We are grateful to the Goddard Space Flight Center team for 
introducing us to the COBE data. 


\end{document}